\documentclass[twocolumn]{article}

\usepackage[utf8]{inputenc}
\usepackage{siunitx}
\usepackage{array,url,kantlipsum}
\usepackage[pdftex]{graphicx}
\newcommand{\Mark}[1]{\textsuperscript{#1}}
\usepackage{hyperref}
\usepackage[super,sort&compress,comma]{natbib} 
\usepackage[left=1.7cm, right=1.7cm, top=1.785cm, bottom=2.0cm]{geometry}
\usepackage[format=plain,justification=justified,singlelinecheck=false,font={stretch=1.125,small,sf},labelfont=bf,labelsep=space]{caption}
\usepackage{balance}
\usepackage{textcomp}
\usepackage{xcolor}

\begin{document}
\sffamily
\twocolumn[{%


\LARGE Direct in- and out-of-plane writing of metals on insulators by electron-beam-enabled, confined electrodeposition with submicrometer feature size\\[1.em]
 
 \large Mirco Nydegger\Mark{1},
        Zhu-Jun Wang\Mark{2,3},
        Marc Willinger\Mark{2,4},
        Ralph Spolenak\Mark{1,}$^{*}$,
    and Alain Reiser\Mark{1,5}\\[0.5em]
 \normalsize
 
\Mark{1}\small{Laboratory for Nanometallurgy, Department of Materials, ETH Zürich, Vladimir-Prelog-Weg 1-5/10, Zürich 8093, Switzerland}\\ 
\Mark{2}\small{Scientific Center of Optical and Electron Microscopy, ScopeM, ETH Zürich, Otto-Stern Weg 3, 8093 Zürich, Switzerland}\\
\Mark{3}\small{School of Physical Science and Technology, ShanghaiTech University, Shanghai, China}\\
\Mark{4}\small{School of Natural Science, Department of Chemistry, Technical University of Munich, Lichtenbergstraße 4, 85747 Garching, Germany}\\
\Mark{5}\small{Department of Materials Science and Engineering, Massachusetts Institute of Technology, MA-02139 Cambridge, USA}\\[0.5em]

  $^{*}$To whom correspondence should be addressed; E-mail: ralph.spolenak@mat.ethz.ch\\[0.5em]

\textbf{Additive microfabrication processes based on localized electroplating enable the one-step deposition of micro-scale metal structures with outstanding performance, e.g. high electrical conductivity and mechanical strength. They are therefore evaluated as an exciting and enabling addition to the existing repertoire of microfabrication technologies. Yet, electrochemical processes are generally restricted to conductive or semiconductive substrates, precluding their application in the manufacturing of functional electric devices where direct deposition onto insulators is often required. Here, we demonstrate the direct, localized electrodeposition of copper on a variety of insulating substrates, namely Al\textsubscript{2}O\textsubscript{3}, glass and flexible polyethylene, enabled by electron-beam-induced reduction in a highly confined liquid electrolyte reservoir. The nanometer-size of the electrolyte reservoir, fed by electrohydrodynamic ejection, enables a minimal feature size on the order of 200~nm. The fact that the transient reservoir is established and stabilized by electrohydrodynamic ejection rather than specialized liquid cells could offer greater flexibility towards deposition on arbitrary substrate geometries and materials. Installed in a low-vacuum scanning electron microscope, the setup further allows for \textit{operando}, nanoscale observation and analysis of the manufacturing process.}\\[1em]
 
}]



\section{Introduction}

Localized electroplating allows the direct and maskless in- and out-of-plane fabrication of metal structures with micro- and nano-scale resolution. Such 3D structures find applications in electronics \cite{Hu2010Meniscus-ConfinedBonds, Schurch2023DirectElectrodeposition}, microswimmers \cite{Ye20203D-PrintedApplications}, microelectromechanical systems \cite{Pagliano2022MicroAccelerometer}, or photonic structures \cite{Richner2016PrintableResolution, Berglund2022AdditiveComponents}. To unlock their full potential when it comes to materials range and quality, such micron-scale additive manufacturing technologies must offer access to a large variety of device-grade metals as well as compatibility with a wide range of substrate materials (especially when targeting functional devices). This combination of requirements is a current challenge for micron-scale additive manufacturing of metals \cite{Hirt2017AdditiveScale, Reiser2020MetalsProperties}. As an advantage, the electrical or structural performance of electrochemically synthesized metals is often excellent in the as-deposited state\cite{ Hirt2017AdditiveScale, Reiser2020MetalsProperties, Schurch2023DirectElectrodeposition} and high-temperature annealing steps can be avoided. Unfortunately, electrochemical AM technologies are limited to conductive and semiconductive substrates that can provide the charges needed for reduction and deposition of metal cations. Compatibility with different substrates is thus low---especially in light of the fact that metals are almost always paired with insulators in functional devices. Notably, this limitation also precludes electrochemical AM to be used for multi-material deposition where one material is an insulator.

Here, we aim at extending the range of substrate materials accessible to electrochemical AM to ready this family of technologies for processing on a wide range of substrates. In general, electrochemical deposition of metals on insulating substrates is unlocked when alternative sources of electrons can bypass the need for electron transfer from the substrate. This can be achieved for example with chemical reduction agents, which are widely employed for electroless plating of metal coatings\cite{Djokic2005ELECTRODEPOSITIONPRACTICE}. A strategy that is better suited for selective, localized deposition is the use of electron beams to initiate electrodeposition\cite{Wang2019Shape-controlledMicroscopy}. Such experiments are typically conducted withing electron microscopes. In previous approaches, the liquid electrolyte was either present in the form of stationary droplets, micro- to millimeters in width \cite{Fisher2015RapidPrecursor, Randolph2013Capsule-freeMicroscope,  Bresin2014LiquidMicroscopy, Esfandiarpour2017FocusedSolutions, Esfandiarpour2021LimitingSurfactants}, or contained in a closed liquid cell \cite{Donev2009Electron-beam-inducedPrecursor,Schardein2011Electron-beam-inducedSolutions,Ocola2012GrowthPrecursor, Bresin2013Site-specificNanostructures, Bresin2013Electron-beam-inducedLiquids, Park2017ElectrochemicalDemand}. However, both approaches for introducing the liquid precursor into an electron microscope are not very well suited for reproducible, high-resolution deposition of 3D structures with a high flexibility to choose different substrates. On the one hand, the use of specialized, electron-transparent liquid cells limits the choice of substrates (so far, polyimide \cite{Schardein2011Electron-beam-inducedSolutions} and Si\textsubscript{3}N\textsubscript{4}\cite{Grogan2014BubbleBeam, DenHeijer2014PatternedBeam} windows have been used), but probably also the widespread applicability because of the technical challenges. On the other hand, methods that utilize an electrolyte reservoir in the form of a droplet on the substrate require less specialized equipment, but are limited in geometric freedom and resolution. Previous approaches relied on \textit{in-situ} hydration of a hygroscopic precursor with low-pressure water vapour \cite{Esfandiarpour2017FocusedSolutions, Bresin2014LiquidMicroscopy}, or injection of the electrolyte through fine capillaries \cite{Fisher2015RapidPrecursor, Randolph2013Capsule-freeMicroscope}. However ,the fact that the electrolyte droplet is stationary and fixed in size limits the maximal height of a printed object, and its variation in thickness can be a challenge for reproducible deposition with high resolution. Both limitations are tied to the mean free path of electrons in a liquid, which is only a couple of \textmu m for typical acceleration voltages of $\leq$ 30~keV\cite{Wiklund2011AWater, Meesungnoen2002Low-energyWater}. Consequently, there is a maximum thickness of the liquid layer above which no deposition on the substrate is possible. It also caps the maximum out-of-plane build space in a single droplet. While a high acceleration voltage offers a larger penetration depth and therefore a larger out-of-plane build space, it can also cause inadvertent increase in minimal feature size due to the increased interaction volume of the primary electrons with the electrolyte (and substrate). For example, it was observed that well-defined writing was restricted to the edges of the liquid droplet where the liquid layer is thin\cite{Esfandiarpour2017FocusedSolutions}. In summary, the previously utilized liquid reservoirs demonstrate a feasibility of electron beam-induced localized electrodeposition but are inflexible regarding compatible substrates, limit the available (out-of-plane) build space and can present challenging environments for constant feature size.

\begin{figure*}[ht]
 \centering
 \includegraphics[scale=1]{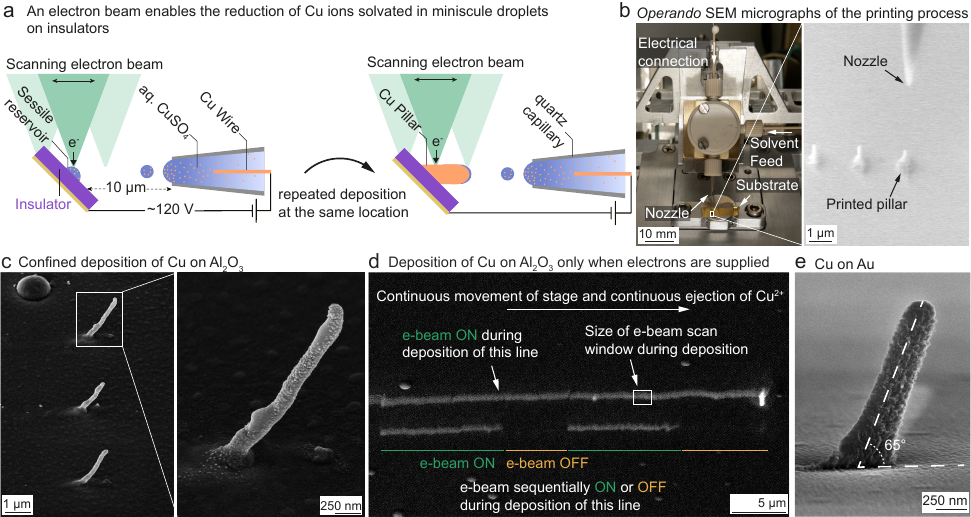}
 \caption{\textbf{: Direct and local electrochemical deposition of Cu on insulators in a scanning electron microscope (SEM).} \textbf{(a)} Schematic of the e-beam-enabled, electrohydrodynamic redox printing process (EHD-RP) for direct electrodeposition of metals on insulating substrates. A sessile, nanoscale reservoir of copper ions is established on the substrate by electrohydrodynamic ejection of ion-loaded solvent droplets from a quartz nozzle positioned approximately 5-10~µm above the substrate. The metal ions are reduced by a scanning e-beam while the solvent evaporates. A voltage applied between a Cu anode and the conductive back of the substrate initiates both, the electrohydrodynamic ejection and the anodic dissolution of the Cu wire. The 45\textdegree  angle of the nozzle to the substrate is due to current space constraints below the SEM pole piece. \textbf{(b)} Left: a photograph of the nozzle holder with solvent feed and sealed copper wire placed above the substrate, which is mounted on a 45\textdegree  tilt stage. Right: \textit{operando} secondary electron (SE) micrograph captured during the printing process. \textbf{(c)} Post-deposition SE micrographs of Cu pillars deposited on Al\textsubscript{2}O\textsubscript{3} (Cu anode in aq. H\textsubscript{2}SO\textsubscript{4} with pH~3, 26~Pa N\textsubscript{2}). The minimal feature size is on the order of 200~nm. The inclination is a result of the 45\textdegree  angle between nozzle and substrate.  Tilt angle of substrate for the image capture: 45\textdegree. \textbf{(d)} The electron beam is inducing reduction of Cu on Al\textsubscript{2}O\textsubscript{3}. The upper line was formed during a continuous lateral stage movement with an unblanked e-beam. Note that the e-beam (20~kV) is scanned in a reduced window. During the deposition of the lower line, the e-beam was blanked periodically (with an otherwise unchanged printing protocol). The temporary blanking of the e-beam during continuous ejection of Cu ions interrupted the deposition of Cu. \textbf{(e)} The same setup allows for deposition on conducting substrates. In this case, the electron beam can be used as a mere \textit{operando} imaging tool. Post-deposition SE micrograph of a Cu pillar printed on a conductive gold substrate. Tilt angle of substrate for the image capture: 85\textdegree.}
  \label{fig_technique}
\end{figure*}

Herein we present an approach to electron beam-enabled, local electrodeposition that circumvents these challenges by the use of a mobile, nanoscale electrolyte reservoir which is formed and maintained on demand by electrohydrodynamic ejection from a printing nozzle. The process combines electrohydrodynamic redox-printing (EHD-RP)\cite{Reiser2019Multi-metalScale} with electron-beam-induced reduction in an environmental (\textit{i.e.} low vacuum) SEM. EHD-RP utilizes electrohydrodynamic ejection of minuscule, ion-loaded solvent droplets towards a substrate. Upon impingement on the substrate, the ions are reduced to a metallic state, while the solvent evaporates (Fig. \ref{fig_technique}a)\cite{Reiser2019Multi-metalScale, Menetrey2022TargetedPrinting, Nydegger2022AdditiveStructures}. The EHD-RP technology (and the electrohydrodynamic printing principle in general) enables direct deposition on a wide range of substrates and substrate geometries (apart from a range of thin-film and bulk substrates, it has been used for deposition onto high-aspect-ratio FIB lift-out grids\cite{Rohner20203DNanowalls}). The mobile electrolyte droplet that travels across the substrate with a relative movement of the printing nozzle guarantees reproducible and constant feature size across large areas\cite{Reiser2019Multi-metalScale}. Finally, as the electrolyte droplet remains sessile atop a growing structure, the deposition of high-aspect-ratio structures is straightforward and wires 170~nm in width with an aspect ratio of 400 have been printed\cite{Reiser2019Multi-metalScale}. These features set the approach apart from previous work on electron-beam-induced electrodeposition. Furthermore, the combination of electrohydrodynamic ejection and electrochemical reduction guarantees the deposition of metals of high density and purity (and hence excellent mechanical and good electrical properties\cite{Reiser2019Multi-metalScale, Reiser2020MetalsProperties, Menetrey2023Microstructure-drivenInterconnects}) at deposition speed that is competitive to ink-based technologies with sub-micrometer resolution. In addition, on-the-fly modulation of either the feed of cations or the printing voltage enables local control of composition, nanoporosity\cite{Reiser2019Multi-metalScale}, or grain size\cite{Menetrey2022TargetedPrinting}, and multi-component electrolytes can be used for the deposition of binary and ternary alloys\cite{Porenta2023Micron-scalePrinting}.

Crucially, the original EHD-RP work relied on electron-transfer from conductive and semiconductive substrates. With the here-presented use of an electron-beam as an alternative source of electrons, we unlock direct deposition onto insulators. We demonstrate here that the combination of electron-beam enabled deposition and EHD-RP facilitates the fabrication of pure and dense Cu structures on different substrates such as conductive Au thin films, insulating alumina and low melting-point polymeric substrates. 

\section{Results and Discussion}

\begin{figure*}[h]
\centering
 \includegraphics[scale=1]{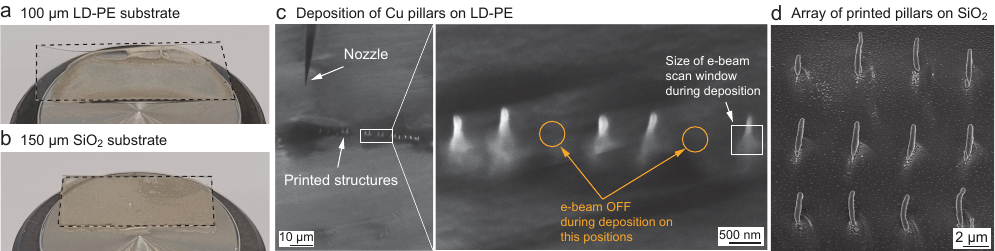}
 \caption{\textbf{: Deposition of Cu onto thick glass and PE:}  Photographs of \textbf{(a)} 100~\textmu m-thick polyethylene foil and \textbf{(b)} 150~\textmu m-thick SiO\textsubscript{2} slide used as substrates for deposition. The substrates are mounted on a metal backing. \textbf{(c)} SE-SEM micrographs of a linear array of Cu pillars printed on low-density polyethylene. The beam was blanked for two array positions, as indicated in the zoomed image. No deposition occurred at these positions. The images were recorded with the environmental SEM directly after deposition. Note that the poor image quality is partially due to charging of the polymer substrate. \textbf{(d)} Post-deposition, ex-situ SE-SEM micrograph of an array of Cu pillars grown directly on glass.}
  \label{fig_substr}
\end{figure*}

\begin{figure*}
\centering
 \includegraphics[scale=1]{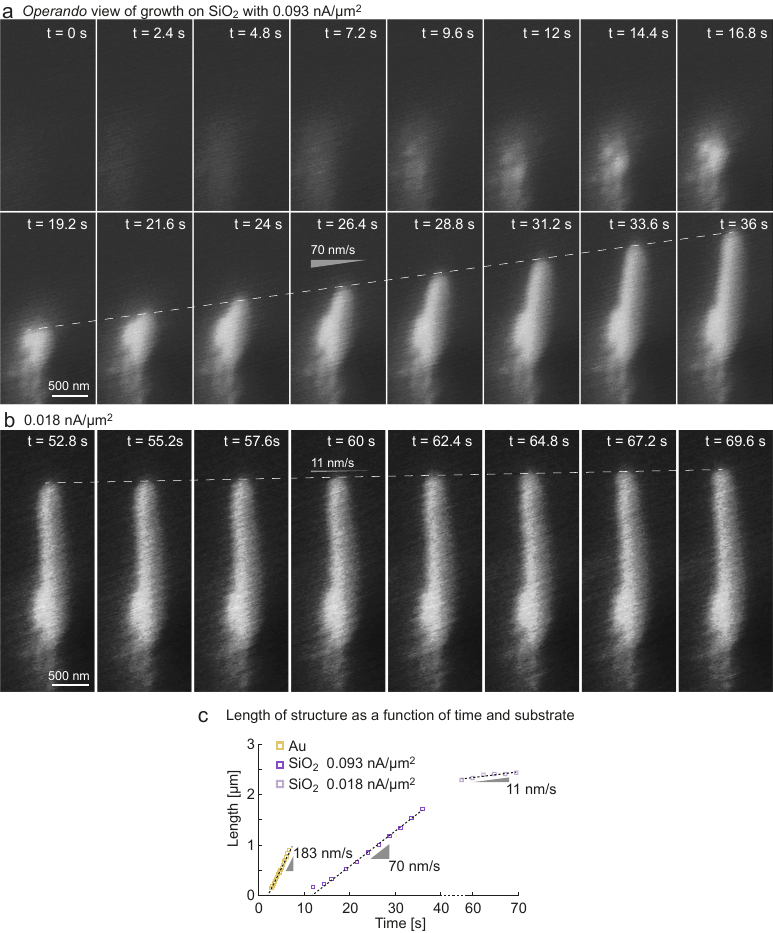}
 \caption{\textbf{: The growth rate scales with the electron beam current density:} \textbf{(a,b)} Sequence of \textit{operando} micrographs documenting the decrease of the growth speed of a Cu pillar upon a reduction of the beam current density. \textbf{(a)} Sequence of the first portion of the pilalr deposited with an average e-beam current density of 0.093~nA/\textmu m\textsuperscript{2} (beam current: 1.2~nA, scan area: 18.64~\textmu m\textsuperscript{2}, time for one scan: 1.2~s). From the image sequence, one can deduce a growth rate of 70~nm/s. \textbf{(b)} A reduction of the average beam current to 0.018~nA/\textmu m\textsuperscript{2} (same beam parameters but a scan area of 97~\textmu m\textsuperscript{2}) slowed the growth speed of the same pillar to just 11~nm/s. In general, note that the diagonal streaks visible in the micrographs likely originated from vibrations. \textbf{(c)} Corresponding plot of the measured length of the pillar as a function of time and current density. For comparison, an \textit{operando} measurement of the growth velocity of a Cu pillar on a Au substrate is shown as well (Supplementary Figure \ref{Supp:GrowthAu}). The higher growth rate on metal substrates suggest that the beam current density was a factor limiting the growth rate on insulators.}
  \label{fig_speed}
\end{figure*}

\subsection{Direct, electron-beam enabled deposition on insulators}

The combination of a nanoscale electrolyte reservoir---as utilized in EHD-RP---with the reducing action of an electron beam is central to the here-presented approach to expand the high-resolution EHD-RP metal printing strategy to insulating substrates. A sessile reservoir of metal ions is established on the substrate by electrohydrodynamic ejection of ion-loaded solvent droplets from a quartz nozzle positioned approximately 5-10~µm above the substrate (Fig. \ref{fig_technique}a). These metal ions are reduced by a scanning e-beam while the solvent evaporates. A voltage of 120-140~V applied between a Cu anode and the conductive back of the substrate initiates both, the electrohydrodynamic ejection and the anodic dissolution of the Cu wire. For easy access to an electron beam as well as \textit{operando} observation of the printing process, we developed an EHD-RP deposition system that can be housed in an environmental scanning electron microscope (SEM) (Fig. \ref{fig_technique}b. For details, see the experimental section and SI Fig. \ref{Supp:setup}). Typical chamber pressures used were 30-40~Pa N\textsubscript{2} or Ar. An inclined substrate facilitates simultaneous access of the e-beam and the printing nozzle. The SEM micrograph in Fig. \ref{fig_technique}b---recorded during the printing process---shows the nozzle above the substrate as well as Cu pillars deposited from an aqueous CuSO\textsubscript{4} solution (1~mM). The choice of the solvent is important. Aqueous electrolytes proved to be compatible with e-beam irradiation and were used for all experiments. In contrast,  with acetonitrile (a standard solvent for EHD-RP), instantaneous clogging of the printing nozzle was observed upon exposure to the electron beam (Fig. \ref{Supp:ACN}). This incompatibility is likely a consequence of the electron-beam induced decomposition of the organic solvent inside of the electron-transparent quartz nozzle\cite{Fisher2015RapidPrecursor}. 

Out-of-plane Cu structures can be grown directly on an insulating film of Al\textsubscript{2}O\textsubscript{3} (1~\textmu m Al\textsubscript{2}O\textsubscript{3} on Au) when the electron beam is irradiating the deposition site (Fig. \ref{fig_technique}d). The inclination of the printed pillars is due to the nozzle not being mounted perpendicular to the substrate (which ensures an out-of-plane growth direction perpendicular to the surface\cite{Reiser2019Multi-metalScale}) but at an angle of 45\textdegree. In principle, the nozzle could also be mounted perpendicular to the substrate, but the angle was chosen because of space constraints below the pole piece. The angle of pillars to the substrate is typically larger than 45\textdegree (60°-65°, measured for five pillars). In the applied electric field, trajectories of ejected droplets probably bend towards the substrate, causing an increase of the impact angle and hence growth direction (simulation in Fig. \ref{Supp:Traj}).

Importantly, e-beam irradiation is a causal and necessary factor for deposition. Even with continuous ejection of Cu ions, blanking of the electron beam interrupts the growth of Cu immediately. This is demonstrated by printing two lines of Cu, once with continuous e-beam irradiation (upper line), once with interrupted e-beam irradiation (lower line in Fig. \ref{fig_technique}d). For both lines, the substrate was translated at a constant speed (V\textsubscript{stage} = 2~\textmu m s \textsuperscript{-1}) and the bias applied between anode and substrate was kept constant (hence, the electrohydrodynamic ejection was constantly operating). The scanning electron beam (acceleration voltage: 20~kV) irradiated a small window at the location where the droplets hit the substrate (droplet impact is usually perceived as a dark spot on the substrate). With continuous e-beam exposure, a continuous line is formed. In contrast, the deposition is interrupted upon intermittently blanking the e-beam at the positions indicated in the lower line (with otherwise unchanged parameters compared to the upper line). It is thus concluded that the electron beam is directly enabling the deposition of Cu on insulating substrates (more details in section 2.2).

Direct deposition onto bulk insulating substrates is as straightforward as on insulating thin films. Deposition on a 150~\textmu m-thick SiO\textsubscript{2} and a 100~\textmu m-thick LD-PE foil is demonstrated in Fig. \ref{fig_substr}. To minimize e-beam-induced damage of the PE, the acceleration voltage had to be lowered to 5~keV. In general, the thickness of the substrate is limited by the electric field that has to be established between the conductive back plate and the anode in the nozzle to initiate electrohydrodynamic ejection. With the maximum voltage of our power source, we were limited to thicknesses <1~mm. While currently not implemented on our system, extractor electrodes (a counter electrode mounted in close proximity to the nozzle) can forgo the need for a conductive electrode below the substrate and have successfully been demonstrated for high-resolution EHD printing\cite{Chaaban2020OnNanoprinting}.

The setup established in this work also allows for deposition on conducting substrates. In this case, the electron beam can be used as a mere \textit{operando} imaging tool (Supplementary Figure \ref{Supp:GrowthAu}b). A Cu pillar printed on a conductive gold substrate is shown in Fig. \ref{fig_technique}e. In general, an advantage of printing in a low-vacuum SEM in comparison to deposition in high-vacuum is that the ionization processes in the gas phase can suppress an e-beam induced charging of isolating substrates in a wider area. It could also open a door towards the direct deposition of oxides, carbides or nitrides by utilizing the formation of reactive species in the gas-phase by the electron beam.

\subsection{Mechanism: e-beam reduction}
Two separate processes are underlying successful deposition by EHD-RP: the electrohydrodynamic ejection of nanometer-sized droplets loaded with metal cations, and the reduction of metal ions striking the substrate. The latter requires a source of electrons, which in previous work has been the conductive substrate. In the present work, we hypothesize that reduction occurs by the primary or secondary beam electrons. To confirm this hypothesis, we will first present how the growth dynamics change with changes in the electron beam current density. Subsequently, we will attempt to rule out alternative mechanisms.

The growth speed of Cu pillars on 150~\textmu m SiO\textsubscript{2} is a function of the e-beam current density. In Fig. \ref{fig_speed}, a series of \textit{operando} SE micrographs (frame time: 1.2~s) illustrates the marked decrease in growth velocity of a single Cu pillar upon the reduction of the current density. The pillar starts nucleating at multiple locations before merging into a single pillar. To avoid an influence of the nucleation phase on the measured growth velocity, the growth dynamics of a single pillar after nucleation was compared under two current densities (instead of comparing growth rates of two pillars). Following the nucleation, the pillar growed linearly with 70~nm/s when irradiated with an averaged current density of 0.093~nA/\textmu m\textsuperscript{2} (Fig. \ref{fig_speed}a). A reduction of the current density to 0.018~nA/\textmu m\textsuperscript{2} (which was achieved by increasing the scanned area by a factor of 5.2) reduced the average growth speed to 11~nm/s (a reduction by a factor of 6.4). The reduction of the growth velocity is roughly similar as the reduction of the current density. It should be noted that the results suggest that the decrease in growth speed is not merely related to the reduction of the dwell time of the beam on the growing pillar. If this was the case, one would expect an acceleration of the growth rate in (a), as the total dwell time of the beam on the pillar increases by a factor of around 3.3 between the bottom left and bottom right image---an increase in growth rate that is not observed. Measuring a diameter of the printed pillar of 320~nm and assuming cylindrical geometry, the growth velocities relate to volumetric deposition rates of 0.0056 \textmu m\textsuperscript{3}/s and 0.00088~\textmu m\textsuperscript{3}/s, respectively. These rates are lower than those observed for deposition on conductive substrates. For comparison, a growth speed of 183~nm/s was measured for the growth of a Cu pillar on a Au thin film using the same concentration of the electrolyte and similar deposition conditions (the series of SEM images is shown in \ref{Supp:GrowthAu}b). During the deposition on Au, the ejection rate is the limiting factor. At a set ejection potential, the same should be the limiting factor during the printing on insulating substrates, given the electron beam current density is high enough to not itself limit the growth rate. As we observed significantly lower growth rates on insulating substrates, and the observed growth rates scaled with the current density, we conclude that in these experiments, the current density was limiting the growth velocity. Interestingly, deposition with the lower low e-beam current lead to the formation of a ring like feature around the pillar (Supplementary Figure \ref{Supp:GrowthAu}a). This resembles a phenomenon known in inkjet printing where charged droplets are deflected from charged dielectric surfaces\cite{Yudistira2010FlightPrinting}. This observation could be indicative of a positive charging of the growing pillar that is not adequately compensated by the low electron current density and further hint towards a rate-limiting nature of the electron beam current density. Overall, we expect that similar growth velocities as achieved on conductive layers can be accessed on insulating substrates by employing higher current densities (e-beam currents many times higher than those tested are easily accessible).

The fact that the growth dynamics are independent of the ejection voltage but scale with the e-beam current density, in combination with the successful deposition on thick substrates, can exclude alternative mechanisms that could sustain the reduction on insulating substrates. Alternative processes that are independent from the e-beam could be: tunneling or leakage currents, an electric breakdown of the insulating material induced by the high applied electric fields. In addition, radiation induced conductivity of the substrate as a beam-dependent mechanism can be added to the list \cite{Shiiyama1998ElectricalIrradiation,Bai2003ElectronFilms}. All of these processes could at least be entertained as possible causes for the observed electrodeposition on the 1~\textmu m thin Al\textsubscript{2}O\textsubscript{3} film. Yet, the deposition on 150~\textmu m SiO\textsubscript{2} convincingly excludes these mechanisms. First, the thick substrate rules out leakage currents potentially caused by pinholes in the thin film, and tunneling currents. Second, the penetration depth of the electron beam in SiO\textsubscript{2} is at maximum a few micrometers, excluding radiation-induced changes of the substrate and potential, accompanying changes in conductivity. And finally, the dielectric breakdown field of glass (E\textsubscript{bd} $>$ 175~V/\textmu m \cite{Fischer2018DielectricGlass}) is well below the field we are operating in. In addition, none of the beam-independent processes can explain the proportionality of the growth rate with the e-beam current density or the interruption of deposition upon blanking of the beam. We therefore conclude that indeed the electron beam causes the reduction of the metal ions, either directly through primary and secondary electrons or through charged species created by ionization processes in the gas phase.

\subsection{Microstructure of
printed Cu}
\begin{figure}[h!]
 \includegraphics[scale=1]{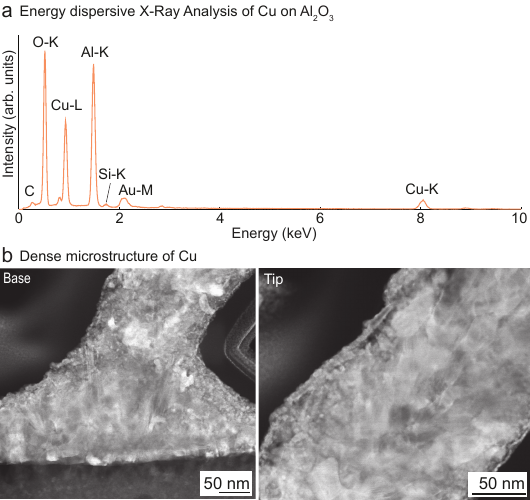}
 \caption{ \textbf{: Pure copper of high density: (a)} An EDX spectrum of a printed Cu pillar on Al\textsubscript{2}O\textsubscript{3} confirms the high purity of the deposits. Carbon is frequently observed in structures printed with EHD-RP, but could also arise as artefacts from observation with an electron beam. The Au, Al and O signals originate from the substrate. \textbf{(b)} Dark field scanning transmission electron micrograph of a Cu pillar on Al\textsubscript{2}O\textsubscript{3}. The material is polycrystalline and of high density, with a homogeneous and pore-free interface to the substrate (HAADF micrographs show density variations that could indicate some porosity, Fig. \ref{Supp:TEM}).
 }
  \label{fig_micro}
\end{figure}

Energy dispersive X-ray spectroscopy analysis confirms that the Cu deposits are of high purity (Fig. \ref{fig_micro}a). Al, O, Au and Si peaks are assumed to be signals from the substrate (however, some oxygen has been detected in previous reports, and we cannot exclude the presence of oxygen in the deposits \cite{Nydegger2022AdditiveStructures}). The absence of sulfur (which could originate from the sulphate anions of the used Cu salt) indicates that no counter-ions are co-deposited. This is in agreement with previous work that has found no counter-ions in the deposits when printing from dilute CuSO\textsubscript{4} and ZnSO\textsubscript{4} solutions ($\leq$ 1~mM)\cite{Porenta2023Micron-scalePrinting}. In comparison to our previous EHD-RP experiments in ambient atmosphere, we typically detected higher amounts of carbon in structures printed within the here-presented setup. It is currently unclear whether the carbon arises from the in-situ observation of the structures with the electron beam itself, or whether the contamination originates from the solvent injection system which has much larger surface area and is more challenging to clean than the single-use nozzles used in previous work.

Transmission electron microscopy of a pillar reveals a highly dense, polycrystalline microstructure (Fig. \ref{fig_micro}b)---a microstructure that is consistent with previous reports for Cu by EHD-RP. Low magnification bright field, dark field and high-angle annular dark field (HAADF) TEM images can be found in \ref{Supp:TEM}. It should be noted that density variations are recorded with the HAADF detector, possibly indicating nanoscale porosity. The interface between pillar and substrate is significantly less porous than previously observed when printing with water \cite{Nydegger2022AdditiveStructures}. Such interface porosity has been attributed to hydrogen evolution\cite{Stevanovi1998HydrogenAlloys}. It is thus possible that the present experimental conditions suppress significant hydrogen evolution during the initial stages of deposition. Interestingly, the base of the pillar---multiple times the width of the pillar itself---is much broader than what has previously been reported. This difference might be due to some positive charging of the substrate and accompanying broadening of the EHD droplet stream. More careful balancing of ion and electron currents should hence be a focus of future experiments.

\section{Conclusions and Outlook}
In conclusion, we report the direct electrochemical fabrication of copper structures on a variety of conductive and non-conductive substrates, enabled by electron-beam-induced reduction. We demonstrated the deposition of Cu structures with minimal feature size of 200~nm, high purity and dense microstructure. The approach does not utilize specialized liquid cells nor large and stationary electrolyte droplets, but rather transient, nanoscale electrolyte reservoirs established by electrohydrodynamic injection. The mobility of these reservoirs significantly enhances the flexibility of the process. It has previously unlocked the growth of structures with very high aspect ratios\cite{Reiser2019Multi-metalScale} or complex 3D geometries that are reproduced with high accuracy when field-guiding of droplets is properly taken into account. Further, we see no fundamental limitation of the build space other than the range of the stages and the size of the vacuum chamber. 
To increase reproducibility, the \textit{operando} observation of the printing process will further help to understand the growth dynamics and optimize path planning and droplet guiding---possibly enabling a simulation-guided deposition model that has been applied successfully to e-beam induced deposition of complex 3D nano-geometries\cite{Menetrey2023NanodropletPrinting}. Important for application, the here demonstrated, direct deposition on insulators proofs compatibility of the EHD-RP process with all classes of materials (deposition on semiconductors was shown earlier\cite{Reiser2019Multi-metalScale}). In concert with recent demonstrations of the growth of semiconductors\cite{Nydegger2022AdditiveStructures} as well as binary and ternary alloys\cite{Porenta2023Micron-scalePrinting} by EHD-RP, the extension to insulating substrates may constitute an important step towards a sub-micrometer, multi-material AM technology for the deposition of functional, multi-material devices.
 
\section{Experimental}
\subsection{Materials}
Nozzles for deposition were fabricated on a P-2000 micropipette puller system (Sutter Instruments) from round quartz capillaries (CM Scientific, Item CV6084-Q-100). Nozzle opening diameters were determined with a Quanta 200F (Thermo Fisher Scientific, former FEI) in low vacuum mode (30~Pa). Nozzles with diameters of 120 \textendash 200~nm were used for printing. Au thin-films were used as conductive substrates. 80~nm thick Au films on top of a 5~nm thick Ti adhesion layer on (100)-Si wafers (SiMat) were prepared in our laboratory sputter facility by DC magnetron sputtering (PVD Products Inc.). For insulating substrates, an additional 1~\textmu m Al\textsubscript{2}O\textsubscript{3} was deposited onto the Au-coated wafers by e-beam evaporation. All wafer substrates were subsequently cut to 0.4~cm × 2.0~cm pieces with a dicing saw. For deposition on insulating and flexible substrates, a LD-PE foil (Minigrip) of 100~\textmu m thickness was used. The foil was mounted on a metal backing with silver glue. Cover slips (150~\textmu m, Menzel, Thermo Fisher Scientific) were used as SiO\textsubscript{2} substrates, glued with silver paste on a SEM stub. Immediately prior to experiments, Cu wires (Alfa Aesar, 0.25~mm diameter, 99.999\% metal basis) were etched in nitric acid ($>$65\%, Sigma Aldrich) for 10~s and rinsed with deionized water. The electrolyte used for deposition was an aqueous solution of 1~mM CuSO\textsubscript{4} (Sigma Aldrich, 99.999\% metal basis) that was prepared with LC/MS grade water (Fisher Chemical, Optima LC/MS). Initial experiments on Au were carried out with a Cu anode in aq. H\textsubscript{2}SO\textsubscript{4} with pH~3.

\subsection{Setup} 
The printing setup was constructed inside a Quattro S SEM (Thermo Fisher Scientific) equipped with a Schottky-type Field Emission Gun. A custom built gas supply system allowed deposition in Ar or N\textsubscript{2} atmosphere in a pressure range from 10\textsuperscript{-4} to 4000~Pa. The in-house built printing system, comprised of the nozzle, anode and solvent feed as well as translation stages for the substrate, was mounted on the standard SEM stage (See supplementary Fig. \ref{Supp:setup}). The nozzle was held in a PEEK T-cross piece  with FEP ferrules (Valco Instruments Co. Inc.) that enabled a gas-tight connection to a PEEK feed line for the electrolyte solution as well as the anode wire (Fig. \ref{fig_technique}c). This T-cross was mounted on a manual stage for coarse positioning of the nozzle. The feed line connected to a glass syringe (Hamilton) outside the SEM chamber. The anode wire was connected to and polarized by an external power source (B2962, Keysight). The substrate was mounted on a three axis (XYZ) piezo nanopositioning system with a range of 200~µm (Nano-3D200 with Nano-drive controller, Mad City Labs). For coarse movements of the substrate, the piezo system was mounted atop a coarse linear stage (M110, Physik Instrumente). The positioning stages and the power source were controlled with a custom Matlab script.

\subsection{Printing procedure}
Prior to experiments, the feed line was flushed with electrolyte solution. The nozzle and the complete fluid system were filled with electrolyte before evacuation of the SEM chamber. Similarly, the nozzle was brought into close proximity of the substrate using the manual approach stage. Then, the chamber was pumped to a typical vacuum level of 30 \textendash 50~Pa (Ar or N\textsubscript{2} atmosphere). Electrohydrodynamic ejection was established by the application of voltages in the range of 120\textendash 140~V between the anode wire and the common ground of the SEM chamber (which connected to the substrate). The DC nature of the applied voltage resulted in a constant deflection of the electron beam that was easily corrected for. The distance of nozzle to substrate was typically 5\textendash 10~\textmu m.
 
\subsection{Analysis} 
High Resolution Scanning electron microscopy (SEM) was performed with a Magellan 400 SEM (Thermo Fisher Scientific, former FEI) equipped with an Octane Super EDXsystem (EDAX, software: TEAM). Tilt angles were 55 or 85$^\circ$. HR-SEM images were taken in immersion mode with an acceleration voltage of 5~kV. 3~nm Pd-Pt were typically sputter-coated onto samples printed on insulators to ensure conductivity for post-deposition SEM analysis. A dual-beam Helios 5UX (Thermo Fisher Scientific) with a focused Ga\textsuperscript{+} liquid metal ion source was used for focused ion beam (FIB) milling of transmission electron microscopy (TEM) crosssections. Prior to FIB-milling the pillar was coated by a protective carbon layer, which is visible in the TEM image as a light grey background. TEM was performed on a Talos F200X (Thermo Fiscer Scientific) operated at 200~kV.

\section{Author Contributions}
A.R. devised the concept, designed and built the setup with guidance from Z.-J.W. and M.W.. A.R., Z.-J.W. and mainly M.N. performed deposition experiments. M.N. provided SEM and FIB analysis. A.R. and M.N. visualised the data. M.N. wrote the original paper draft. All authors discussed the results and reviewed the manuscript.

\section{Conflicts of interest}
The Authors declare no conflicts of interest.

\section{Acknowledgement}
This work was funded by Grant no. SNF 190826 and partially supported by Grant no. SNF 200021-188491. We thank M. Menétrey and N. Porenta for experimental support and the simulations. Electron-microscopy analysis was performed at ScopeM, the microscopy platform of ETH Zürich. E-beam evaporation was done by the Laboratory Support group of ETH Zurich, D-PHYS.

\bibliography{references} 
\bibliographystyle{rsc} 

\newpage
\newcommand{\beginsupplement}{%
        \setcounter{table}{0}
        \renewcommand{\thetable}{S\arabic{table}}%
        \setcounter{figure}{0}
        \renewcommand{\thefigure}{S\arabic{figure}}%
     }

\setcounter{page}{1}
\beginsupplement
\onecolumn
\noindent\Large{\textbf{Supplementary Information}}\\
\\
\\
\noindent\LARGE{\textbf{Direct in- and out-of-plane writing of metals on insulators by electron-beam-enabled, confined electrodeposition with submicrometer feature size}} \\
\noindent\large
\\
Mirco Nydegger, Zhu-Jun Wang, Marc Willinger, Ralph Spolenak, Alain Reiser
\normalsize 

\vspace{2cm}

\begin{figure*}[ht!]
   \centering
   \includegraphics[scale=1]{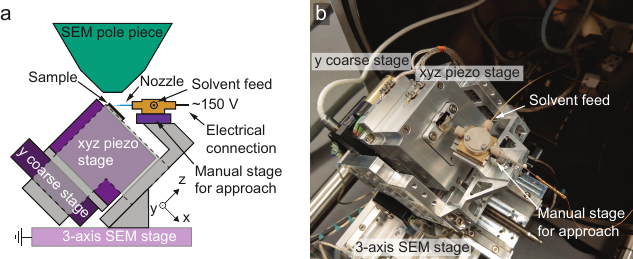}
   \caption{\textbf{: Setup: (a)} Schematic side-view of the built setup. The SEM stage is used to bring the nozzle into focus under the SEM pole-piece. The piezo stages enable the movement of the sample relative to the nozzle. The nozzle has to be brought into the range of the z-axis piezo stage by a manual approach stage. \textbf{(b)} Fotograph of the built setup. The PEEK cross that connects the nozzle, the sacrificial anode and the solvent feed can be seen.}
\label{Supp:setup}
\end{figure*}

\begin{figure*}[ht!]
   \centering
   \includegraphics[scale=1]{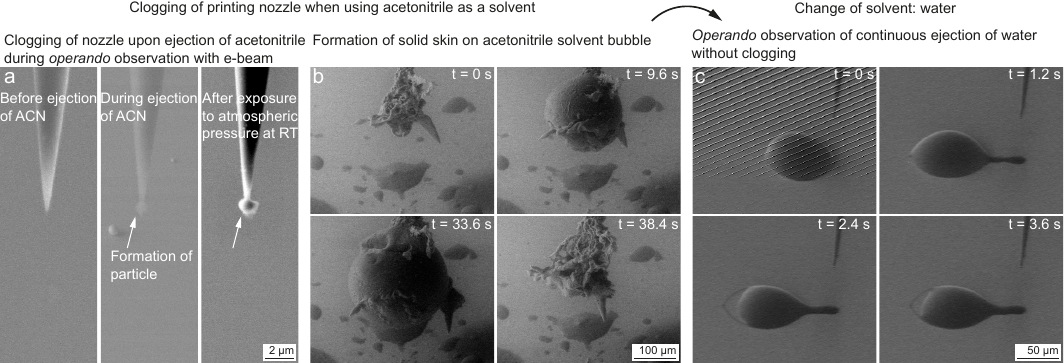}
   \caption{\textbf{Acetonitrile as a solvent results in potential e-beam induced polymerization.} \textbf{(a)}, With acetonitrile, the solvent previously used for EHD-RP\cite{Reiser2019Multi-metalScale}, immediate clogging of the nozzle during imaging with the e-beam was encountered. Left: nozzle before ejection of acetonitrile. Middle: \textit{operando} SEM image of the nozzle during attempted ejection of acetonitrile. The formation of a particle at the nozzle apex is obvious. Right: SEM image of the same nozzle after venting and pumping the chamber. The presence of the particle after exposure to atmospheric pressure confirms that it is not frozen acetonitrile. \textbf{(b)} \textit{operando} observation of the injection of acetonitrile from a nozzle with an opening of several micrometers (thus the large solvent volume). A solvent bubble forms, apparently supported by a skin around the growing bubble. Upon rupture of the skin, the solvent evaporates and only the skin is retained. The behavior in (a) and (b) hints towards e-beam polymerization of the solvent. \textbf{(c)} \textit{operando} observation of the ejection of pure water with a large nozzle at a very high flow rate (chamber pressure: 38 Pa). The use of water as a solvent avoids issues with e-beam-initiated polymerization of the solvent.}
\label{Supp:ACN}
\end{figure*}

\begin{figure*}[ht!]
   \centering
   \includegraphics[scale=1]{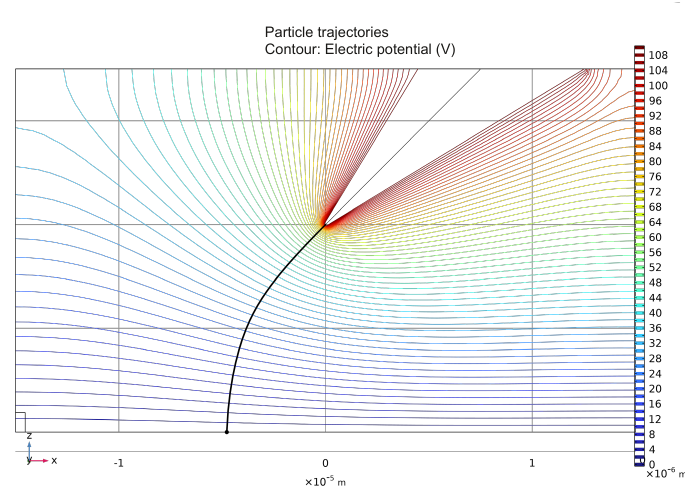}
   \caption{\textbf{: Simulation of droplet trajectory:} A Comsol simulation of a point charge moving through an electric field. As in the printing experiments, a planar electrode and a nozzle inclined with 45° are used in the simulation. The colored contours indicate the electric field strength. The trajectory of the charged particle bends towards the substrate, impacting at an angle greater than 45°. This behavior supports our observation of an inclination angle of printed pillars less than 45°. Used parameters: substrate at 0V, V set at the capillary surface at  110V. The model calculates the trajectory for a point charge subjected to an electrostatic and a drag force assuming r = 95 nm and charge = 358 e. \cite{Menetrey2023NanodropletPrinting}.}
\label{Supp:Traj}
\end{figure*}

\begin{figure*}[ht!]
   \centering
   \includegraphics[scale=1]{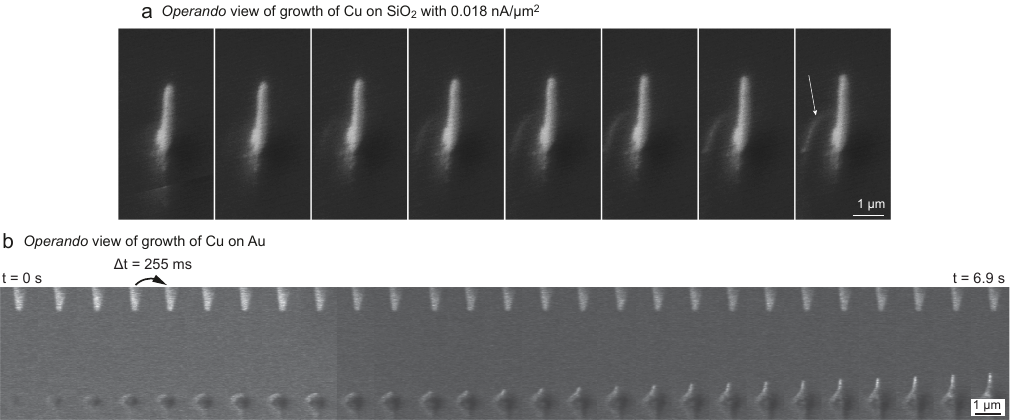}
   \caption{\textbf{Additional \textit{operando} observations of a Cu pillar printed on SiO\textsubscript{2} and Au} \textbf{(a)} low magnification sequence of SE micrographs shown in \ref{fig_speed}b. During the slow growth, the emergence of a ring around the pillar can be observed.  \textbf{(b)} Sequence of \textit{in operando} SE micrographs reveiling the growth dynamics of a Cu Pillar on a Au substrate. The acquisition time of a single image was 255 ms. The growth rate of the Cu pillar is faster than what has been observed on insulating substrate (with similar beam currents), presumably because the beam current density is not a limiting factor (electrons for reduction are supplied through the substrate and not only through the electron beam).}
\label{Supp:GrowthAu}
\end{figure*}

\begin{figure*}[ht!]
   \centering
   \includegraphics[scale=1]{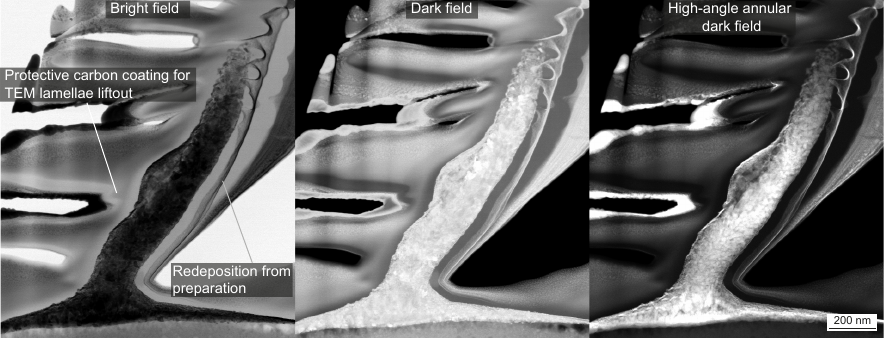}
   \caption{\textbf{Low magnification TEM images:}  Bright field, dark field and high angle annular dark field (HAADF) images of a polished Cu pillar. A homogeneous microstructure is observed along the pillar axis without major porosity. The top of the pillar appears to be thinner. We attribute this to the preparation. The protective carbon coating was not dense, thus creating the contrast-rich features on the left side of the images.}
\label{Supp:TEM}
\end{figure*}



\end{document}